# Condenser Pressure Influence on Ideal Steam Rankine Power Vapor Cycle using the Python Extension Package Cantera for Thermodynamics

**Osama A. Marzouk**

College of Engineering, University of Buraimi, Sultanate of Oman
osama.m@uob.edu.om (corresponding author)

ABSTRACT

**This study investigates the Rankine vapor power thermodynamic cycle using steam/water as the working fluid, which is common in commercial power plants for power generation as the source of the rotary shaft power needed to drive electric generators. The four-process cycle version, which comprises a water pump section, a boiler/superheater section, a steam turbine section, and a condenser section, was considered. The performance of this thermodynamic power cycle depends on several design parameters. This study varied a single independent variable, the absolute pressure of the condenser, by a factor of 256, from 0.78125 to 200 kPa. The peak pressure and peak temperature in the cycle were fixed at 50 bar (5,000 kPa) and 600°C, respectively, corresponding to a base case with a base value for the condenser's absolute pressure of 12.5 kPa (0.125 bar). The analysis was performed using the thermodynamics software package Cantera as an extension of the Python programming language. The results suggest that over the range of condenser pressures examined, a logarithmic function can be deployed to describe the dependence of input heat, the net output work, and cycle efficiency on the absolute pressure of the condenser. Each of these three performance metrics decreases as the absolute pressure of the condenser increases. However, a power function is a better choice to describe how the steam dryness (steam quality) at the end of the turbine section increases as the absolute pressure of the condenser rises.**

## I. INTRODUCTION

In thermodynamic power cycles or heat engines, a set of processes are sequentially arranged to allow the conversion of heat into work, particularly mechanical shaft rotation. The vapor power cycle is a subcategory of thermodynamic power cycles in which the working fluid used in the processes changes its phase between liquid and vapor (gas), being a two-phase cycle [1-5]. The Rankine cycle is a popular example of a vapor power cycle. In this cycle, the source of heat can be combustion that occurs externally (outside the main cycle itself), renewable solar heat, waste heat (hot exhaust gases from a separate gas turbine power cycle, resulting in a "combined cycle"), or heat released from a nuclear reactor. The working fluid can be water and steam or another organic substance [6-18]. An important application of the Rankine cycle is in electric power generation in commercial power plants, which is essential to provide electricity to residential and commercial consumers [19-24]. The performance of the steam Rankine cycle depends on the various parameters adopted in the cycle design. These parameters include the peak temperature occurring at the inlet of the steam turbine [25] and the minimum pressure occurring at the condenser [26]. In [27], the influences of the peak pressure (boiler pressure) and the peak temperature (turbine inlet temperature) on the performance of an ideal four-stage Steam Rankine Cycle (SRC) were investigated as two independent design variables. However, the influence of the condenser pressure was not studied. This study can be viewed as a continuation of [27] considering also the effect of the condenser pressure, which is the minimum pressure that occurs during the Rankine cycle.

The simplified ideal Rankine cycle consists of a pump section, a boiler section with an included superheater, a condensing-type steam turbine section, and a condenser section [28]. "Ideal" refers to ignoring losses in the turbine and pump section, assuming that they operate isentropically [29-35]. As the process within the condenser is a phase change of water at a constant temperature that is implied by pressure, the minimum pressure within the condenser directly specifies the condensation temperature within it [36], which is the minimum temperature in the four-stage SRC. The peak pressure (uniform boiler pressure), the peak temperature (superheating temperature at the turbine inlet), and the lowest pressure (uniform condenser pressure) are the three main conditions needed to fully specify a unique ideal four-stage SRC.

In [37], the effect of the condenser conditions in the Rankine cycle was studied, but not for steam as the working fluid. This study considered the Organic Rankine Cycle (ORC), in which the working fluid is an organic fluid, such as refrigerants or hydrocarbons [38-43]. Similarly, in [44], the optimal condenser temperature was investigated for ORC rather than SRC. Although the principles of thermodynamics and the Carnot efficiency law indicate that the cycle efficiency of a Rankine cycle increases as the condenser temperature and thus the condenser pressure decrease due to the drop in the heat-rejection temperature [45-48], this study examines the general profile of such an increase and recommends a regression model. This study also covers the response of the steam quality (dryness) at the steam exit, which is an important factor in the operation of a steam turbine. Furthermore, this study provides a quantitative example of the agreement between three different sources of water/steam properties.

As this study considers a specific condition for superheated steam, the results and regression models developed are limited by this condition. The current study demonstrates a general functional relationship between the condenser pressure as a control variable and the multiple response variables of the SRC, knowing that the particular values and regression constants should change if the superheated steam condition changes. In addition, the modeled SRC is an ideal simplified one compared to the realistic cycles in steam power plants. However, this should not be viewed as a drawback, as this simplicity is aligned with reduced uncertainty because of the smaller number of parameters involved, other than the condenser pressure, which need to be specified for fully simulating the cycle performance. For example, adding a reheater and a second turbine makes the model more realistic, but the model also becomes more complicated and the influence of the condenser pressure becomes harder to recognize with the presence of the intermediate pressure as a new parameter [49-50].

## II. RESEARCH METHOD

This study takes advantage of numerical modeling and Computer-Aided Design (CAD) to simulate the performance of the ideal four-stage SRC [51-63]. A base case is used to establish a reference operational configuration. Figure 1 shows the thermodynamic cycle and its four stages (four processes or four sections). The figure utilizes the $T$-$s$ thermodynamic diagram ($T$: temperature, $s$: specific entropy or entropy per unit mass) with the mound-shaped saturation curve for water. Numbers 1, 2, 3, and 4 correspond to an arbitrary case of this cycle, and numbers 1', 2', 3, and 4' correspond to another arbitrary case of the cycle with the condenser pressure reduced compared to the 1-2-3-4 cycle. The process 1-2 (or 1'-2') represents isentropic (constant entropy) compression in the water pump section. The process 2-3 (or 2'-3) encompasses pump isobaric (constant pressure) heat addition (pre-heating, then boiling, and finally superheating) in the water boiler pump section. Process 3-4 (or 3-4') represents an isentropic expansion in the steam turbine section. Finally, the process 4-1 (or 4'-1') represents isobaric-isothermal condensation (constant pressure and temperature) in the condenser section. The condensate at state 4 (or 4') is in the form of saturated liquid water, thus all vapor steam has transformed into liquid water, and it is ready for compression again to such an extent that the cycle is repeated over and over in a closed loop formed by the four states 1-2-3-4 (or 1'-2'-3-4').

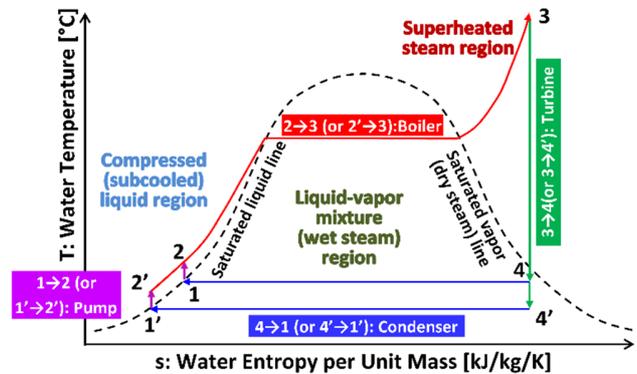

Fig. 1.　Illustration of the ideal four-stage Steam Rankine Cycle (SRC).

This figure displays the effect of reducing the condenser pressure. States 1, 2, and 4 are altered accordingly to the new states 1', 2', and 4'. However, state 3 (end of superheating and entrance of the steam turbine section) is unchanged. State 3 remains at the base value for all the condenser pressures examined. Table I summarizes the base case values for the ideal four-stage SRC modeled here. This table also lists the range of absolute condenser pressures covered. There are nine condenser pressure values (from 0.78125 kPa or 0.0078125 bar to 200 kPa or 2 bar) [64]. These pressure values were carefully selected to form a geometric sequence, with each value (starting from the second) being twice the previous [65]. The use of such a geometric sequence (rather than having an equal spacing between values) is explained by the wide range of the condenser pressure values covered, spanning more than two orders of magnitude. Therefore, the geometric sequence allows similar emphasis on the lower part and the upper part of that range.

TABLE I.　MODELING PARAMETERS

| Property | Value |
|---|---|
| Absolute boiler pressure ($p_2$, $p_a$) | 50 bar (5 MPa) |
| Superheated steam temperature ($T_a$) | 600 °C (873.15 K) |
| Isentropic efficiency of the pump | 100% |
| Isentropic efficiency of the turbine | 100% |
| Gross loss factor, to express the power plant efficiency ($a$) | 90% |
| Range of covered absolute condenser pressure ($p_4$) | 0.78125 kPa (0.0078125 bar) – 200 kPa (2 bar) |
| Exact values of the covered absolute condenser pressure ($p_4$) | 0.78125 kPa, 1.5625 kPa, 6.25 kPa, 12.5 kPa (the base value), 25 kPa, 50 kPa, 100 kPa, 200 kPa |

The isentropic efficiency for compression in the pump section and expansion in the turbine section are set at their ideal value of 100% [66]. Therefore, the cycle is described as "ideal". This eliminates additional uncertainty because any "non-ideal" values would interfere with the aim of the study, which is to quantify the influence of only one design variable (condenser pressure) with reduced influence from other parameters. Despite this, an overall loss factor of 90% is assumed, which absorbs the neglected losses in the pump section and turbine section. This gross loss factor is used to convert the net power from the ideal Rankine thermodynamic cycle into an electric output from the hypothetical power plant that utilizes such a cycle. The specific system-level gross loss factor is close to the efficiency of a large efficient steam turbine [67-68]. The former can be set to another value by the reader, without impacting the core ideal Rankine cycle itself. This is an advantage, due to the flexibility and ease of representing the effect of losses, such as friction in moving parts, heat transfer to the surroundings, and electric generator loss [69-72].

If the enthalpy per unit mass of water (specific enthalpy) at the $i^{th}$ state is ($h_i$), the work transfer per unit mass of water is $w$, and the heat transfer per unit mass of water is $q$, then the following equations describe the mathematical model of the four processes of the ideal Rankine cycle. These equations are derived from the first law of thermodynamics, for the case of an open system (control volume) with a steady flow [73-75].

pump: $w_p = h_2 - h_1$         (1)

boiler: $q_b = h_3 - h_2$         (2)

turbine: $w_t = h_3 - h_4$         (3)

condenser: $q_c = h_4 - h_1$         (4)

net cycle work per unit mass: $w_{net} = w_t - w_p$     (5)

cycle efficiency: $\eta_{cyc} = w_{net}/q_b = 1 - q_c/q_b$     (6)

electric work per unit mass: $w_{elec} = \alpha \, w_{net}$     (7)

power plant efficiency: $\eta_{pp} = \alpha \, \eta_{cyc} = w_{elec}/q_b$     (8)

where $q_b$ represents the input heat per unit mass for the entire thermodynamic cycle. The condition of isentropic compression in the pump section is satisfied by enforcing equal entropy per unit mass (or specific entropy) before and after the compression, thus in state 1 and state 2:

$s_2 = s_1$         (9)

Similarly, the condition of isentropic expansion in the turbine section is satisfied by enforcing equal entropy per unit mass before and after the expansion, thus in state 3 and state 4:

$s_4 = s_3$         (10)

The computational simulations were conducted using the thermodynamics software package "Cantera" (version 2.6.0) as an add-on library to the Python programming language (version 3.8.8) [76-79]. All needed thermodynamic properties of the water (as a liquid, a vapor, or as a mixture of both) were accessed directly through the Cantera object "Water" as a pure fluid with its equation of state based on published properties [80-81]. Figure 2 demonstrates the first 11 lines of the Python code for the analysis. In lines 7-11, the simulation parameters are specified and can be edited. Among these parameters, only *p_min* (absolute condenser pressure) is changed from one simulation case to another. In this figure, the base value of the absolute condenser pressure (12.5 kPa) appears. Figure 3 exhibits the last 10 lines of the output results from the simulation code, which are in the form of formatted text data.

Like the previous figure, this figure also corresponds to the base value of the absolute condenser pressure (12.5 kPa).

```python
1  import cantera as ct
2  print('Cantera version is ', ct.__version__)
3  import sys
4  print("Python version is ", sys.version)
5
6  # input parameters
7  eta_pump = 1.      # pump isentropic efficiency
8  eta_turbine = 1.   # turbine isentropic efficiency
9  p_max = 50.0e5     # maximum pressure (Pa)
10 p_min = 12.5e3     # minimum pressure (Pa)
11 T_max = 873.15     # maximum temperature (K)
```

Fig. 2.     Illustration of Python/Cantera code for the simulations.

```
**************** Cycle Analysis *************
Input pump work per kg (kJ/kg) = 5.043
Output turbine work per kg (kJ/kg) = 1336.989

Net output work per kg (kJ/kg) = 1331.946
Input heat per kg (kJ/kg) = 3450.927

Heat rejected per kg (kJ/kg) = 2118.981

Cycle efficiency = 38.597%
```

Fig. 3.     Illustration of the output returned by the Python/Cantera code.

This study selected four performance metrics to describe their response to the change in the absolute condenser pressure. These were:

- Specific input heat (input heat per unit mass of water) to the boiler ($q_b$), in MJ/kg (megajoule of heat per kg of water)

- Specific net output work (net output work per unit mass of water) from the thermodynamic cycle ($w_{net}$), in MJ/kg. In addition, the estimated electric power output per unit mass ($w_{elec}$), in MJ/kg, was also studied.

- Thermodynamic cycle efficiency ($\eta_{cyc}$), as a percentage. Moreover, the estimated power plant efficiency ($\eta_{pp}$) was also examined.

- Steam quality (dryness fraction) at the exit of the steam turbine section and the inlet of the condenser section ($x_4$), as a percentage. The steam quality (or dryness fraction, or dryness) is the mass fraction of water vapor in a mixture of water vapor and liquid water [82-84]. The opposite (or complement) of steam dryness is steam wetness, which is the mass fraction of liquid water in a mixture of water vapor and liquid water [85-86].

A regression model was developed for each one of the four performance metrics to achieve good alignment with the nine training data points (for each metric) without showing spurious oscillation due to overfitting (successfully matching the training points but failing to maintain a smooth profile that can be applied to other data points) [87-88]. For example, a high-order polynomial may achieve a low value of deviations from the nine training data points, but with erratic behavior between the training data points. The goodness-of-fit for the regression models is judged by the $R^2$ value, which is the coefficient of determination [89-90]. Although the $R^2$ value has a specific interpretation in linear regression models [91], it is still used here for nonlinear regression models as a measure of the closeness of the predicted to the training values [92].

### III. VALIDATION

The simulation results depend largely on the water's thermodynamic properties used in the simulation. As a validation step for the method adopted in Cantera for reporting these properties, the base case of the Rankine cycle was simulated engaging two other independent sources of water properties, and the results were compared with those obtained deploying Cantera for 10 cycles. Table II illustrates this comparison. As mentioned above, $x_4$ is the steam quality (or dryness fraction) of the liquid-vapor water mixture leaving the turbine [93-94]. The steam quality is desired to be high, as near as possible to its upper limit of 100%. A low dryness fraction means that small water droplets (fine mist) impact the turbine blades at a high speed, which causes large stresses in these blades, leading to erosion [95-96]. A small fraction of liquid droplets can be tolerated, so that the steam dryness fraction does not decrease below about 90% [97].

One of the other two sources of thermodynamic water properties used in the benchmarking is the online calculator freely provided by the large UK company "Spirax Sarco Limited", which offers products and services in the field of steam and its applications [98]. The second source of thermodynamic water properties utilized in the benchmarking is the miniREFPROP free desktop software by the National Institute of Standards and Technology (NIST) of the U.S.A. [99]. It is a small version of the larger commercial software program REFPROP by NIST, which is also called the NIST Standard Reference Database 23 (SRD 23) [100-102]. For water, the equation of state in REFPROP is based on the 1995 formulation of the International Association for the Properties of Water and Steam (IAPWS) [103-106].

TABLE II.    BENCHMARKING OF BASE CYCLE RESULTS

| Quantity | Spirax Sarco | NIST miniREFPROP | Cantera (here) |
|---|---|---|---|
| $T_1$ [°C] | 50.2403 | 50.240 | 50.288 |
| $h_1$ [kJ/kg] | 209.835 | 210.35 | 210.5134 |
| $s_1$ [kJ/kg/K] | 0.7070 | 0.70692 | 0.707436 |
| $T_2$ [°C] | 50.4163 | 50.420 | 50.470 |
| $h_2$ [kJ/kg] | 214.864 | 215.39 | 215.5568 |
| $h_3$ [kJ/kg] | 3,666.330 | 3,666.8 | 3,666.4839 |
| $s_3$ [kJ/kg/K] | 7.2606 | 7.2605 | 7.258869 |
| $x_4$ [%] | 88.9975% | 88.9975% | 88.9556% |
| $h_4$ [kJ/kg] | 2,329.195 | 2,329.69 | 2,329.4947 |
| $\eta_{cyc}$ [%] | 38.5955% | 38.5950% | 38.597% |

For the Cantera results in the benchmarking table, it should be noted that the enthalpy per unit of mass (specific enthalpy) and the entropy per unit of mass (specific entropy) listed are the raw values reported by Cantera after subtracting the corresponding values at the triple point of water (liquid phase) at 0.01°C and 0.610 kPa [107-109]. This was necessary to have values comparable to what is reported by the Spirax Sarco web tool and by the NIST miniREFPROP offline tool. These reference triple point values were -1,5970.805415 kJ/kg for specific enthalpy, and 3.519982 kJ/kg/K for specific entropy.

This change in values does not affect the computations, because it is actually the change in these specific properties that matters, rather than their individual values [110-112].

When comparing the values in the benchmarking table, it becomes evident that the results are consistent. For example, the base-case cycle efficiencies are exactly the same up to four significant digits (38.60%). In addition, the base-case steam dryness fractions at the turbine exit are identical in the three benchmarking values if displayed with four significant digits (88.90%). This favorable agreement supports the validity of the Cantera simulation method, which is applied in all the results presented in the next section.

### IV. RESULTS

This section presents the response curves for the four performance metrics of the ideal four-stage SRC as the absolute pressure in the condenser changes. Due to the large difference between the smallest and the largest values of the condenser pressure data points studied, a logarithmic scale (with base 2) is used for the horizontal axis representing the absolute condenser pressure (kPa). The axis settings were adjusted so that the axis labels and vertical major lines exactly coincide with the nine values of the condenser pressures examined.

A regression model was developed for each metric, employing the Microsoft Excel built-in trendline tool. Linear, exponential, and polynomial options were excluded because they were unable to match the profile of the training data points. The two remaining qualitatively acceptable fitting types were the logarithmic function (general form: $y = a \ln(x) + b$) and the power function (general form: $y = x^a$), where $x$ refers to the absolute condenser pressure ($p_4$) in kPa, $y$ refers to the particular performance metric being modeled, and $a$ and $b$ refer to optimized model parameters. The fitting with the lowest $R^2$ value is the one recommended here. For accuracy, at least five significant digits are kept in the model parameters provided. Figure 4 demonstrates the variation of the specific input heat (to the boiler section, including the integrated superheater section) with the absolute condenser pressure.

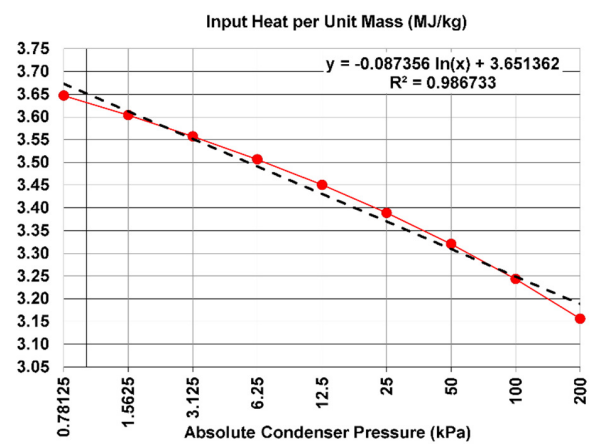

Fig. 4.    Variation of input heat per unit mass with condenser pressure.

The recommended regression model is a logarithmic function as follows:

$$q_b [\text{MJ/kg}] = -0.087356 \ln(p_4[\text{kPa}]) + 3.651362 \quad (11)$$

Heating requirements decline as the condenser pressure increases. While this appears as a desired feature, it should not be judged in isolation from the corresponding net output shaft work of the cycle, whose response curve (per unit mass) is observed in Figure 5. Its recommended regression model is also a logarithmic function, as follows:

$$w_{net}[\text{MJ/kg}] = -0.13474 \ln(p_4[\text{kPa}]) + 1.64988 \quad (12)$$

Similar to the demanded heat, the useful mechanical shaft work declines as the condenser pressure increases. This is an undesirable feature.

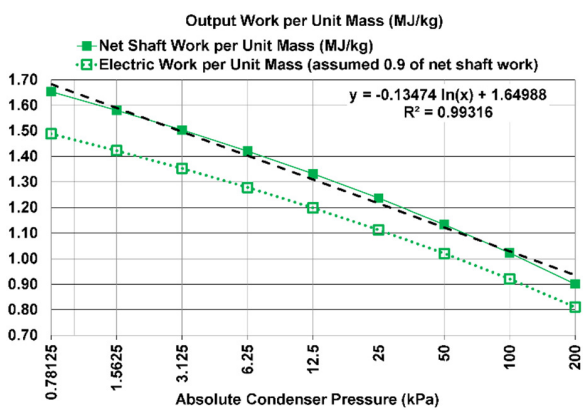

Fig. 5. Variation of net work per unit mass with condenser pressure.

Figure 6 prese the response curve for the cycle efficiency to also decline as the condenser pressure increases. This means that the decline in the net output work is faster than the decline in the required input heat. The response curve is monotonic. Therefore, a lower condenser pressure should be sought as possible for better performance. The theoretical limit for the absolute condenser pressure is the triple point pressure for water (0.610 kPa), where the solid phase of water (ice) begins to appear, and liquid water cannot exist as a stable phase below that pressure [113-114]. This theoretical lower limit on the condenser pressure is superseded by another practical limit that arises from the heat exchange process in the condenser in the case that external cooling water is used to reject the heat contained in the steam being condensed. In this case, the condenser pressure is limited by the saturation pressure of that external cooling water, whose temperature should be lower than the temperature of the condensing steam [115]. For example, if the external cooling water is at 20°C, and a temperature difference of 10°C is desired for proper heat transfer from the condensing steam to the external cooling water [116], then the condenser temperature (the temperature of the steam being condensed) should be 30°C. The corresponding saturation pressure (condensation pressure) at this temperature of 30°C for water is 4.24 kPa [117-118].

The recommended regression model for the cycle's thermodynamic efficiency is also a logarithmic function, with the following form:

$$\eta_{cyc}[-] = -0.029895 \ln(p_4[\text{kPa}]) + 0.454749 \quad (13a)$$

$$\eta_{cyc}[\%] = -2.9895\% \ln(p_4[\text{kPa}]) + 45.4749 \quad (13b)$$

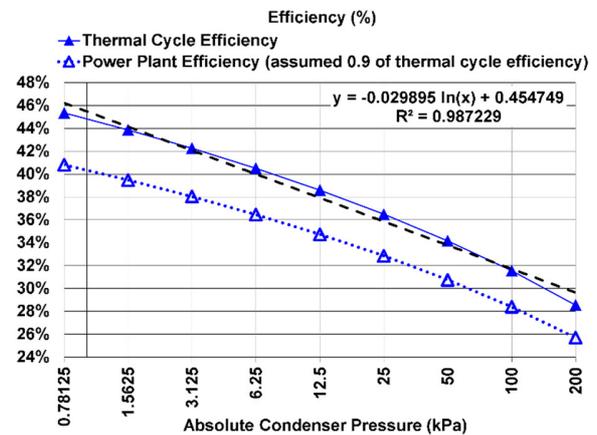

Fig. 6. Variation of efficiency with condenser pressure.

The fourth and final performance metric for the Rankine cycle is the steam dryness fraction after the expansion in the turbine section is completed, and thus the steam (either in a wet form as a mixture of gaseous vapor and small liquid droplets carried with it, or as a purely gaseous phase) is to be sent to the condenser to recover the water as a liquid medium that can be pumped in a new cycle. Figure 7 portrays the response curve for this dryness fraction.

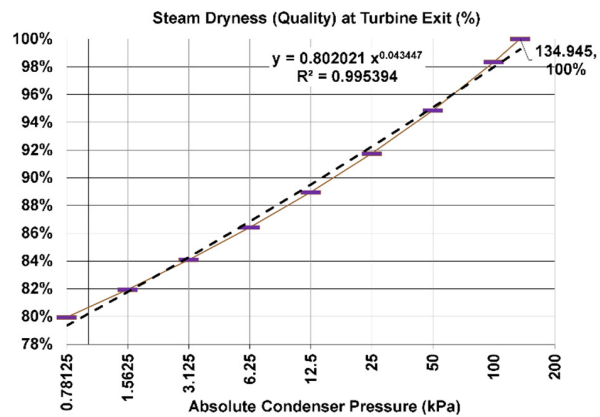

Fig. 7. Variation of steam dryness at turbine exit with condenser pressure.

Unlike the other three performance metrics, the dryness fraction increases as the condenser pressure increases. The dryness fraction increases from 79.95% at an absolute condenser pressure of 0.78125 kPa (which is close to the triple point) to 98.34% at 100 kPa (near the normal atmospheric pressure). The steam exiting the turbine becomes saturated vapor (single phase, but on the verge of partial condensation to change into a two-phase medium) at an absolute condenser pressure of 134.945 kPa. Beyond this pressure, the expanded steam exits the turbine section in a slightly superheated state, so it can be expanded slightly further without developing any liquid fraction. This occurs, for example, at the highest absolute condenser pressure value covered here, which is 200 kPa. The recommended regression model for the dryness fraction of the expanded steam is a power function with the following form:

$$x_4[-] = 0.802021 \, p_4[kPa]^{0.043447} \quad (14a)$$

$$x_4[\%] = 80.2021\% \, p_4[kPa]^{0.043447} \quad (14b)$$

The $R^2$ value for this power regression model is 0.995394. A logarithmic regression model was found to give a lower value of 0.990295 for $R^2$, so it was not selected despite not being very different in its profile from that of the power function model.

Figure 7 indicates that although decreasing the condenser pressure improves the energy conversion efficiency of the cycle (from heat to work), fatigue and stresses can cause problems in the late segments of the steam turbine due to the decline in the steam dryness fraction [119-120]. Therefore, the condenser pressure to achieve higher cycle efficiency should be accompanied by an adaptation in the cycle to limit the formation of liquid fraction in the expanded steam. This, for example, can be achieved by increasing the superheating temperature to the limit that the turbine material can withstand based on the metallurgical properties of the turbine blades [121]. Such a change (increasing the superheating temperature) should shift the location of state 4 in the T-s diagram to the right, closer to the saturated vapor line, where the steam dryness fraction is exactly 100%.

## V. CONCLUSIONS

The condenser pressure of a representative ideal four-stage SRC was varied over a wide range, from an absolute value of 0.78125 kPa (close to the triple point of water, the theoretical lower limit) to 200 kPa (two times the normal atmospheric pressure). The response curves for four performance metrics of this vapor power thermodynamic cycle were computed and nonlinear regression models were developed for each of them. The superheated steam temperature was fixed at 600°C (873.15 K or 1,112 F), and the absolute pressure of the superheated steam was fixed at 50 bar (5 MPa or 725 psi). The thermodynamics software package Cantera within the Python environment was used in the analysis.

Reducing the condenser pressure always improves the cycle efficiency, with a logarithmic dependence on the absolute condenser pressure. Simultaneously, the heat requirements and the net output work decrease as the condenser pressure increases. However, the decrease in net output work is faster than the decrease in the demanded heat, which means that lowering the condenser pressure is a beneficial way to improve cycle efficiency. However, the minimum possible condenser pressure is practically limited in the case of heat exchange with external cooling water. Although decreasing the condenser pressure is advantageous in terms of energy conversion, the steam expands to a lower pressure in the turbine. Therefore, a larger fraction of liquid water is formed in the wet steam (vapor-liquid mixture) near the steam turbine exit, which is structurally disadvantageous for the turbine blades. The steam dryness at the turbine exit depends on the absolute condenser pressure according to a power function.

Although principles of thermodynamics imply that reducing the condenser pressure in the Rankine cycle should result in improving the energy conversion efficiency, this study identifies a mathematical relationship between the cycle effectiveness and the condenser pressure as a controllable design variable for the representative configuration of the ideal four-stage SRC. Furthermore, the relationships between this design variable and other performance metrics of that power cycle were visualized and regression models were developed. Through these quantifiable regression models, a designer of a steam Rankine cycle may make proper decisions during the preliminary design stage concerning how low should the condenser pressure be, by balancing the expected gains from reducing this pressure against any incurred costs or complications due to operating near a vacuum. For example, this study shows that the cycle efficiency can be improved by 10% if the steam is allowed to expand to a normal temperature of 30°C (near ambient levels) but with reduced pressure at about 1/25 of the normal atmospheric level of 100 kPa. Therefore, this study provides a quantitative assessment of role of the condenser pressure in the operation of a Rankine steam cycle. Moreover, this study provided validation for three different sources of the thermodynamic properties of water, manifesting the consistency between them.